\documentclass[aps, twocolumn,prb,preprintnumbers,amsmath,amssymb,superscriptaddress]{revtex4}
\usepackage{amsmath,amsfonts,amssymb,graphics,graphicx,epsfig,color,times,indentfirst,layout,hyperref,subfigure,multirow}
\usepackage{hyperref}
\usepackage{srcltx}
\usepackage{lipsum}
\usepackage{ulem}
\hypersetup{linktocpage}
\usepackage{threeparttable}

\begin{document}

\title{Entanglement negativity in free-fermion systems: An overlap matrix approach}

\author{Po-Yao Chang}
\affiliation{Center for Materials Theory, Rutgers University, Piscataway, New Jersey, 08854, USA}
\affiliation{Department of Physics, University of Illinois at Urbana-Champaign, Urbana, Illinois, 61801, USA}

\author{Xueda Wen}
\affiliation{Department of Physics, University of Illinois at Urbana-Champaign, Urbana, Illinois, 61801, USA}

\date{\today}

\begin{abstract}
In this paper, we calculate 
the entanglement negativity in free-fermion systems by use of the overlap matrices.
 For a tripartite system, if the ground state 
 can be factored into triples of modes, 
 we show that the partially transposed reduced density matrix can be factorized
 and the entanglement negativity has a simple form.
 However, the factorability of the ground state in a tripartite system does not hold in general.
 In this situation, the partially transposed reduced density matrix can be expressed in terms of the Kronecker product 
 of matrices. 
 We explicitly compute the entanglement negativity for the Su-Schrieffer-Heeger model,  the integer Quantum Hall state, 
 and a homogeneous one-dimensional chain.
 We find that the entanglement negativity for the integer quantum Hall states shows an area law behavior.
 For the entanglement negativity of two adjacent intervals in a homogeneous one-dimensional gas, we find agreement with the conformal field theory.
Our method provides a numerically feasible way to study the entanglement negativity in free-fermion systems.
\end{abstract}

\maketitle

\noindent

\section{Introduction}
The study of quantum entanglement 
provides a powerful tool to analyze the many-body states  in condensed matter physics\cite{Kitaev, Levin, Calabrese0, Amico, Eisert,Vidal0,Calabrese00,Calabrese000,Benatti,Benatti2}.
The most celebrated measure of entanglement is given by
the entanglement entropy (von Neumann entropy) $S_A$ of
the reduced density matrix of the subsystem $A$.
This quantity measures the entanglement between two complementary subsystems $A$ and $B$,
 when the total system is in a pure state.
One remarkably common result is an area law behavior of the entanglement entropy in gapped systems,
which grows proportionally with the boundary between two subsystems\cite{Eisert}.
An important exception is the one-dimensional systems at criticality,
where the entanglement entropy has a logarithmic scaling behavior\cite{Calabrese0,Vidal0,Eisert} and
have been understood by use of a conformal field theory (CFT) approach\cite{Calabrese000,Calabrese0}.

However, using the entanglement entropy as a probe to diagonalize the entanglement for a mixed state
is a difficult task. It requires a heavy optimization process.
Hence, an alternative measure of entanglement for a mixed state has been proposed---the entanglement negativity\cite{Eisert1,Vidal,Plenio}.
The advantage of using entanglement negativity is that it only requires linear algebraic computations.
The entanglement negativity is defined as follows.
Suppose a tripartite system is divided into $A_1$, $A_2$, and $B$ subsystems.
The ground state after tracing out the degrees of freedom in a $B$ subsystem is a general mixed state.
The entanglement negativity, which quantifies the entanglement between $A_1$ and $A_2$,
is obtained by partially
 transposing (with respect to the degrees of freedom in $A_2$) the reduced
density matrix $\rho_{A_1 \cup A_2}$, which is denoted as $\rho_{A_1 \cup A_2}^{T_{A_2}}$.
Then the entanglement negativity between subsystems $A_1$ and $A_2$ is defined as
\begin{align}
\mathcal{E} := \ln \mathrm{Tr} |\rho_{A_1\cup A_2}^{T_{A_2}}|,
\end{align}
where the trace norm $|\rho_{A_1\cup A_2}^{T_{A_2}}|$ represents the sum of all absolute values of the eigenvalues of $\rho_{A_1\cup A_2}^{T_{A_2}}$.
Recently, the entanglement negativity has
been extensively studied in numerous many-body systems, e.g.,
harmonic oscillators in one dimension\cite{Audenaert,Marcovitch,Ferraro,Cavalcanti,Anders,Anders1}, 
quantum spin chains\cite{Wichterich2, Wichterich3, Bayat, Bayat2, Bayat3, Santos},  
systems with topological orders \cite{Castelnovo, Lee}, and (2+1) dimensional Chern-Simons theories\cite{EN_CS}.
A CFT approach\cite{Calabrese,Calabrese2} and many useful numerical methods are also developed, including 
tree tensor techniques\cite{Calabrese5},
Monte Carlo simulations\cite{Alba, Chung}, and rational interpolations\cite{Nobili}.
Further studies on non-equilibrium situations\cite{Eisler,Coser,Hoogeveen,Wen} and
finite temperature\cite{Calabrese4, Eisler} refine our understanding of the entanglement negativity.

In this work, we demonstrate a systematical way to compute the entanglement negativity in free-fermion systems
by using the overlap matrices. For the prior works on entanglement negativity in free fermion systems\cite{Eisler2, Coser2, Coser3, Eisler3, Coser4}, 
the quantities people calculated are the moments of partially transposed reduced denity matrix $\left(\rho_{A_1 \cup A_2}^{T_{A_2}}\right)^n$, with $n\ge 2$. 
In other words, the entanglement negativity, which is obtained from
the trace norm $\left|\rho_{A_1 \cup A_2}^{T_{A_2}}\right |$, 
has not been computed in the prior works due to reasons of technical difficulty. 
In our work, we will give an explicit calculation of $\left|\rho_{A_1 \cup A_2}^{T_{A_2}}\right |$ by using the overlap matrices, 
and therefore the entanglement negativity is obtained. 
The overlap matrices are built up from the single-particle states $\phi_n$ 
corresponding to the lowest $N$ (occupied) energy levels of the system\cite{Israel,Calabrese6,Calabrese3,Calabrese7,Ossipov,Drut}:
\begin{align}
[M_\sigma]_{nm} =\int_{\sigma} d^d x \phi_n^*(x) \phi_m(x), \quad, n,m=1,\cdots,N,
\end{align}
where the integration is restricted to the subsystems $\sigma=A_1, A_2$ or $B$ in arbitrary dimension $d$.
We show that for a tripartite system, when the overlap matrices $M_{A_1}$, $M_{A_2}$, and $M_B$
can be simultaneously diagonalized, the ground state of fermions can be factored into triples of modes.
In this situation, the partially transposed reduced density matrix can be factorized and the entanglement negativity
can be obtained from the spectra of the overlap matrices. 
On the other hand, when the simultaneous diagonalizability of the overlap matrices does not hold,
the factorability of the ground state in a tripartite system is failed.
In this situation, the partially transposed reduced density matrix can only be expressed in terms of
the Kronecker product of matrices. Then the entanglement negativity is directly obtained from the partially transposed reduced density matrix.
The overlap matrix approach provides a systematical way to calculate the entanglement negativity
in any free-fermion systems, which can be either gapped topological phases
or critical systems. 
This approach is also applicable for disordered systems and can be extended to higher dimensions.

The paper is organized as follows: 
In Sec. \ref{Sec: EE_pure},
we review the entanglement negativity for a pure state. 
In Sec. \ref{Sec: EE_mix_diag},
we obtain the entanglement negativity for a mixed state when 
the ground state can be factored into triples of modes in a tripartite system.
We find that the entanglement negativity for the integer quantum Hall state satisfies an area law. 
In Sec. \ref{Sec: EE_mix_nondiag}, we derive
the entanglement negativity for a general mixed state.
We compute the entanglement negativity for free fermions on a ring and show the entanglement negativity is in agreement with CFT. 
In Sec. \ref{Sec: conclusion}, we conclude our results.

\section{Entanglement negativity in free-fermion systems for pure states}
\label{Sec: EE_pure}

Let us start with the Hamiltonian for free fermions
\begin{align}
H=\sum_{\alpha=1}^{M} \epsilon_\alpha d^\dagger_{\alpha} d_{\alpha},  \rm{with}  \quad
\epsilon_1 \leqslant \epsilon_2 \leqslant \cdots \leqslant \epsilon_M,
\end{align}
where $d_{\alpha}$ is the fermion operator with corresponding energy $\epsilon_\alpha$,
and $M$ is the number of bands.
The ground state is 
\begin{align}
|\Psi  \rangle=  \prod_{\alpha=1}^{N}  d^\dagger_{\alpha}|0 \rangle, 
\label{Eq: GS}
\end{align}
where $N$ is the total number of particles in the system.
We consider a bipartite system that the system is partitioned into $A$ and $B$ parts
and the single-particle wave function can be partitioned by the projection operator $\mathcal{P}_{A(B)}$. 
Following Ref. [\onlinecite{Israel}],
we introduce an overlap matrix,
\begin{align}
[M_{A (B)}]_{\alpha \beta}&= \langle \mathcal{P}_{A(B)}  u_{\alpha},     \mathcal{P}_{A(B)}  u_{\beta} \rangle \notag \\
&= \int_{r\in A(B)} d^d r \phi^*_\alpha ({\bf r})  \phi_\beta ({\bf r}), \rm{with} \quad 1\leqslant\alpha, \beta\leqslant N,
\label{Eq: overlap_matrix}
\end{align}
where $| u_\alpha\rangle =  d^{\dagger}_{\alpha} | 0 \rangle$ 
and $ \mathcal{P}_{A(B)}$ is the orthogonal projection operator on Hilbert space $A(B)$,
such that  $\mathcal{P}_{A(B)} |u_{\alpha} \rangle = \phi_\alpha ({\bf r})|_{\bf r \in A(B)}$.
Since $M_B=\mathbb{I}-M_A$,
one can find a unitary matrix $U_{i\alpha}$ that simultaneously diagonalizes $M_A$ and $M_B$,
such that $U M_A U^{\dagger}={\rm diag} (P_i)$ and 
$U M_B U^{\dagger}={\rm diag} (1-P_i)$.
The single-particle Hilbert space is decomposed into two parts
$\mathcal{H}=\mathcal{H}_A \bigoplus \mathcal{H}_B$.
The orthonormal modes at $\mathcal{H}_A $ and $\mathcal{H}_B$
are
\begin{align}
|A_i \rangle = \frac{\sum_\alpha U^\dagger_{i \alpha } \mathcal{P}_A | u_\alpha \rangle}{\sqrt{P_i}}, \quad
|B_i \rangle = \frac{\sum_\alpha U^\dagger_{i \alpha } \mathcal{P}_B | u_\alpha \rangle}{\sqrt{1-P_i}}.
\end{align}
Let $E=\rm{span}(|u_\alpha  \rangle; 1\leqslant \alpha \leqslant N)$
is the subspace of the single particle Hilbert space $\mathcal{H}$.
Since $U$ is unitary, we have $E= {\rm span}(U^\dagger_{i\alpha}|u_\alpha \rangle)$.
The ground state in Eq. (\ref{Eq: GS}) can be rotated by this unitary matrix $U$.
By using
\begin{align}
\sum_\alpha U^{\dagger}_{i \alpha}d^\dagger_\alpha | 0 \rangle &= \sum_\alpha U^\dagger_{i\alpha}|u_\alpha\rangle =\sum_\alpha U^\dagger_{i\alpha}(\mathcal{P}_A |u_\alpha\rangle
+\mathcal{P}_B |u_\alpha\rangle) \notag\\
&= \sqrt{P_i} |A_i \rangle + \sqrt{1-P_i} |B_i \rangle,
\end{align}
the ground state under the rotation can be written as (up to a global phase)
\begin{align}
|\Psi  \rangle=  \prod_{i=1}^{N} (\sqrt{P_i} d^\dagger_{Ai} + \sqrt{1-P_i}d^\dagger_{B_i})|0 \rangle, 
\end{align}
where $d^\dagger_{A(B)i}  | 0 \rangle  = |A(B)_i \rangle $.




We can compute the entanglement negativity directly from
the partially transposed density matrix $\rho^{T_{B}}$.
The density matrix from the ground state under occupation basis in Eq. (\ref{Eq: GS}) is
\begin{align}
\rho = \bigotimes_{i}\left(\begin{array}{cc}P_i & \sqrt{P_i(1- P_i)} \\ \sqrt{P_i(1-P_i)} & 1-P_i\end{array}\right),
\end{align}
where the basis is $\{ | 1_{Ai}  0_{Bi}\rangle,  | 0_{Ai}  1_{Bi}\rangle \}$.

The partially transposed density matrix $\rho^{T_{B}}$ is 
\begin{align}
\rho^{T_B} &= \bigotimes_{i} \rho_i^{T_B}  \notag\\
&= \bigotimes_{i} \left(\begin{array}{cccc}P_i & 0 & 0 & 0 \\0 & 1- P_i & 0 & 0 \\0 & 0 & 0 & \sqrt{P_i(1-P_i)} \\0 & 0 & \sqrt{P_i(1- P_i)} & 0\end{array}\right),
\end{align}
where the basis is $\{ | 1_{Ai}  0_{Bi} \rangle,  | 0_{Ai}  1_{Bi}\rangle,  | 0_{Ai}  0_{Bi}\rangle,  | 1_{Ai}  1_{Bi}\rangle \}$.

Notice that the eigenvalues of the Kronecker product of two matrices $W_A \otimes W_B$ are
$\lambda_i \mu_j$, where $\lambda_i$ and $\mu_j$ are the eigenvalues of $W_A$ and $W_B$, respectively.
The eigenvalues of $\rho^{T_B}$ are the product of one of the eigenvalues of each $ \rho_i^{T_B}$.
The set of eigenvalues of $\rho_i^{T_B}$ is $\{\Xi _{i, \alpha} \}=\{ P_i, 1- P_i,  \sqrt{P_i (1- P_i)}, - \sqrt{P_i (1- P_i)} \}$
with $\alpha$ being the label of the elements in the set. 
Hence the entanglement is
\begin{align}
\mathcal{E} &= \mathrm{ln Tr} | \rho^{T_B}| =\mathrm{ln} \prod_{i} \sum_{\alpha} ( |\Xi _{i, \alpha}| ) \notag\\
&= \sum_{i} \mathrm{ln} (1 + 2 \sqrt{P_i (1-P_i)}).
\label{Eq: pure}
\end{align} 
Examples of the entanglement negativity for pure states are shown in App. \ref{App: EE_pure}.

\section{Entanglement negativity in free-fermion systems for mixed states}

\subsection{The ground state in a tripartite system can be factored into triples of modes}
\label{Sec: EE_mix_diag}

Now we generalize the bipartite case to a tripartite system.
The overlap matrices $M_{A_1}$, $M_{A_2}$, and $M_B$ 
are constructed from Eq. (\ref{Eq: overlap_matrix}), with 
the projected single-particle wave function  $\mathcal{P}_{A_1}|u_\alpha \rangle$,
$\mathcal{P}_{A_2}|u_\alpha \rangle$, and $\mathcal{P}_{B}|u_\alpha \rangle$.
If the overlap matrices can be simultaneously diagonalized by a unitary matrix $U_{i \alpha}$, 
we can find the orthonormal modes at the subsystems 
\begin{align}
&|A_{1i} \rangle = \frac{\sum_\alpha U^\dagger_{i \alpha } \mathcal{P}_{A_1} | u_\alpha \rangle}{\sqrt{P_{1i}}}, \quad
|A_{2i} \rangle = \frac{\sum_\alpha U^\dagger_{i \alpha } \mathcal{P}_{A_2} | u_\alpha \rangle}{\sqrt{P_{2i}}}, \\ \notag
&|B_i \rangle = \frac{\sum_\alpha U^\dagger_{i \alpha } \mathcal{P}_B | u_\alpha \rangle}{\sqrt{1-P_{1i}-P_{2i}}}.
\end{align} 
Since the single-particle Hilbert space does note change under unitary transformation,
 i.e., $E={\rm span} ( |u_{\alpha} \rangle)={\rm span} (U^\dagger_{i \alpha } |u_{\alpha} \rangle)$, and
\begin{align}
&\sum_\alpha U_{i\alpha}^\dagger d^{\dagger}_{\alpha} | 0\rangle = \sum_\alpha U_{i\alpha}^\dagger | u_{\alpha} \rangle  \notag \\
&=\sum_\alpha U^\dagger_{i\alpha}(\mathcal{P}_{A_1} |u_\alpha\rangle+\mathcal{P}_{A_2} |u_\alpha\rangle
+\mathcal{P}_B |u_\alpha\rangle)    \notag \\
&= (\sqrt{P_{1i}} |A_{1i}\rangle+
\sqrt{P_{2i}} |A_{2i}\rangle+ \sqrt{1-P_{1i}-P_{2i}}|B_i\rangle ),
\end{align}
the ground state can be factored into triples 
of modes
\begin{align}
|\Psi \rangle &= \prod_i (\sqrt{P_{1i}} d^\dagger_{A_1 i} + \sqrt{P_{2i}} d^\dagger_{A_2 i} +\sqrt{1-P_{1i}-P_{2i}} d^{\dagger}_{B i}) | 0 \rangle.
\label{Eq:gwf1}
\end{align}
where $d^{\dagger}_{\sigma i}|0 \rangle = |\sigma_{i} \rangle $ with $\sigma=A_1, A_2, B$.

The reduced density matrix after tracing part $B$ is
\begin{align}
\rho_A = \bigotimes_{i} \left(\begin{array}{cccc}P_{1i} & \sqrt{P_{1i}P_{2i}} & 0 & 0 \\ \sqrt{P_{1i}P_{2i}}  &  P_{2i} & 0 & 0 \\ 0 & 0 & 1-P_{1i}-P_{2i} & 0 \\ 0 & 0 & 0& 0\end{array}\right),
\label{reduced}
\end{align}
where the basis of the reduced density is $\{ | 1_{A_1} 0_{A_2} \rangle_{i},  |0_{A_1}1_{A_2}\rangle_{i},   |0_{A_1}0_{A_2}\rangle_{i},  |1_{A_1}1_{A_2}\rangle_{i} \}$.

The partially transposed reduced density matrix after transposing the $A_2$ is 
\begin{align}
\rho_A^{T_{A_2}} &= \bigotimes_{i}\rho_{Ai}^{T_{A_2}}  \notag \\
&=\bigotimes_{i}
 \left(\begin{array}{cccc}P_{1i} &0 & 0 & 0 \\ 0  &  P_{2i} & 0 & 0 \\ 0 & 0 & 1-P_{1i}-P_{2i} &  \sqrt{P_{1i}P_{2i}}  \\ 0 & 0 &  \sqrt{P_{1i}P_{2i}} & 0\end{array}\right).
\label{reducedT}
\end{align}
The set of eigenvalues of $\rho_{Ai}^{T_{A_2}}$ is $\{\Xi _{i, \alpha} \}=\{ P_{1i},  P_{2 i}, \frac{1}{2} (1- P_{1i}-P_{2i}  \pm \sqrt{(1-P_{1i}-P_{2i})^2+4 P_{1i}P_{2i}}) \}$
with $\alpha$ being the label of the elements in the set. 
The entanglement negativity is 
\begin{align}
\mathcal{E} &= \mathrm{ln Tr} | \rho_A^{T_{A_2}}  | 
= \sum_{i} \mathrm{ln} (\sum_{\alpha}|\Xi_{i,\alpha}|) \notag\\
&= \sum_{i} \ln (P_{1i} + P_{2i} + \sqrt{(1-P_{1i}-P_{2i})^2+4P_{1i} P_{2i}}).
\label{Eq: EN for mixed state}
\end{align}

Notice that if the probability for all the eigenstates in subsystem $B$ is zero, which can be realized by taking the dimension of Hilbert space of $B$ to be zero,
the entanglement negativity becomes the case of pure states,
\begin{align}
\lim_{\mathrm{dim. } \mathcal{B} \to 0} \mathcal{E} &= \sum_{i} \ln (1 + 2 \sqrt{P_{1i} P_{2i}}) \notag\\
&= \sum_{i} \ln (1 + 2 \sqrt{P_{1i} (1-P_{1i})}),
\end{align}
where $\mathcal{B}$ is the Hibert space of region $B$.

In the following, we consider the entanglement negativity between $A_1$ and $A_2$ regions of
two examples in which the overlap matrices can be simultaneously diagonalized.
The first one is the fully dimerized Su-Schrieffer-Heeger (SSH) model and 
the second one is the integer quantum Hall state (IQHS).

\subsubsection{Dimerized Su-Schrieffer-Heeger (SSH) model}
\label{Sec:SSH}
\begin{figure}[htbp]
\centering
 \includegraphics[height=2. cm] {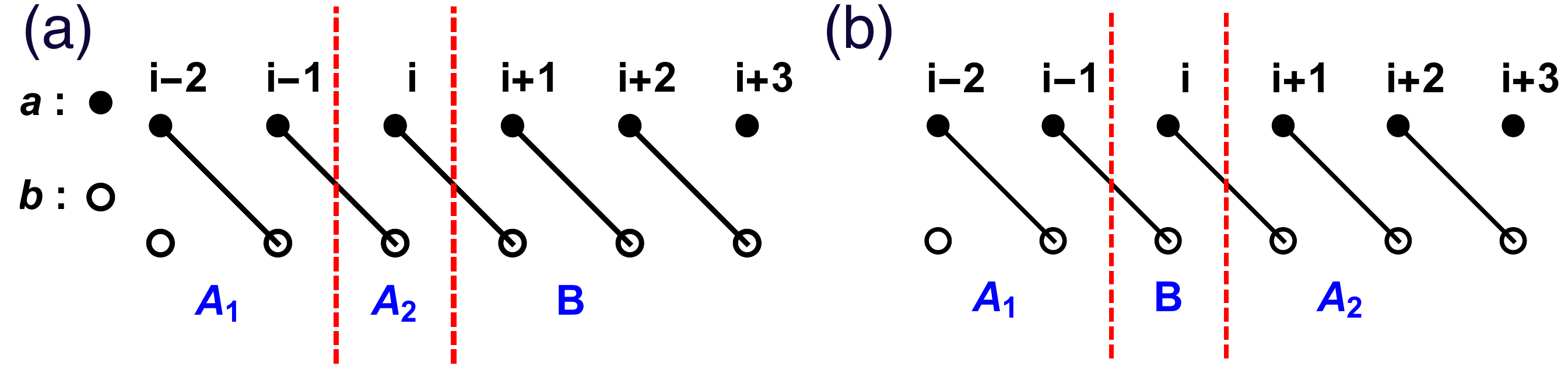}
\caption{ The tripartite configuration of a dimerized Su-Schrieffer-Heeger model.
 The hopping amplitude of solid lines is $t_2$. 
One unit cell contains two orbitals that are pictured by $\bullet$ and $\circ$.
The system is divided 
into $A_1$, $A_2$ and $B$ parts by the red dashed lines.
(a) Adjacent intervals for $A_1$ and $A_2$ regions.
(b) Disjointed intervals for $A_1$ and $A_2$ regions.}
\label{fig:1D_chain}
\end{figure}

The Su-Schrieffer-Heeger model is first introduced in Ref. [\onlinecite{Su79}],
which describes one-dimensional polyacetylene. 
The Hamiltonian of a dimerized SSH model is 
\begin{align}
H=\sum_i t  c^\dagger_{a_{i+1}} c_{b_{i}} + \mathrm{h.c.},
\end{align}
where $t$ is the hopping amplitude and $c_{a(b)_i}$ is the fermion operator. 
The ground state for this dimerized SSH model is
\begin{align}
|\Psi  \rangle = \prod_{\alpha} d^{\dagger}_{\alpha} |0 \rangle,
\end{align}
where $d_{1}^{\dagger} = \frac{1}{\sqrt{2}} (c^{\dagger}_{a, 1}-c^{\dagger}_{b, 2}), d_{2}^{\dagger} = \frac{1}{\sqrt{2}} (c^{\dagger}_{a, 2}-c^{\dagger}_{b, 3}), \cdots,
d_{j}^{\dagger} = \frac{1}{\sqrt{2}} (c^{\dagger}_{a, j}-c^{\dagger}_{b, j+1}), \cdots$. 

We consider an infinite chain
that the system is divided into $A_1$, $A_2$, and $B$ as shown in Fig.~\ref{fig:1D_chain}.
The state $d_{j}^{\dagger} | 0 \rangle=|u_j \rangle$ is 
localized at $a$ orbital on $j$-th site and at $b$ orbital on $(j+1)$-th site.
This localized state is the eigenstate of three overlap matrices, $M_{A_1}$, $M_{A_2}$, and $M_{B}$.
For the geometry shown in Fig. \ref{fig:1D_chain}(a), the corresponding eigenvalues of $M_{A_1}$ and $M_{A_2}$ are, 
\begin{align}
P_{1j}=\langle u_j |M_{A_1}| u_j \rangle 
= \left\{
 \begin{aligned} &1, \quad &j \leqslant i-2,   \\
 &0.5, \quad &j = i-1,   \\
 & 0, \quad  &j\geqslant i.
 \end{aligned}
 \right. 
 \label{eq: A1}   
 \end{align}
  
\begin{align} 
 P_{2j}=\langle u_j |M_{A_2}| u_j \rangle 
= \left\{
 \begin{aligned}
 &0.5, \quad & i-1 \leqslant j \leqslant i,   \\
 & 0, \quad  &{\rm otherwise}.
 \end{aligned}
 \right.
 \label{eq: A2}
\end{align}

For the geometry shown in Fig.\ref{fig:1D_chain}(b), the corresponding eigenvalues of $M_{A_1}$ and $M_{A_2}$ are, 
\begin{align}
P_{1j}=\langle u_j |M_{A_1}| u_j \rangle 
= \left\{
 \begin{aligned} &1, \quad &j \leqslant i-2,   \\
 &0.5, \quad &j = i-1,   \\
 & 0, \quad  &j\geqslant i.
 \end{aligned}
 \right. 
 \label{eq: AA1}   
 \end{align}
  
\begin{align} 
 P_{2j}=\langle u_j |M_{A_2}| u_j \rangle 
= \left\{
 \begin{aligned} &0, \quad &j \leqslant i-1,   \\
 &0.5, \quad &j = i,   \\
 & 1, \quad  &j\geqslant i+1.
 \end{aligned}
 \right.
 \label{eq: AA2}
\end{align}

\paragraph{Adjacent intervals}
From Eqs. (\ref{Eq: EN for mixed state}), (\ref{eq: A1}), (\ref{eq: A2}), 
the entanglement negativity between $A_1$ and $A_2$ (see Fig. \ref{fig:1D_chain}(a))
after tracing out $B$ is
\begin{align}
\mathcal{E} &=\sum_{j} \mathrm{ln} (1-P_{2j} +\sqrt{P_{2j}^2+4P_{1j}(1-P_{1j}-P_{2j})}) \notag\\
&= \ln (0.5 + 0.5 + \sqrt{4\times(0.5)\times(0.5)}) = \ln 2.
\end{align}
This $\ln 2$ comes from the contribution
of the states $|u_{i-1}\rangle$ localized on the boundary between $A_1$ and $A_2$.

\paragraph{Disjointed intervals}
From Eqs. (\ref{Eq: EN for mixed state}), (\ref{eq: AA1}), and (\ref{eq: AA2}),
the entanglement negativity between $A_1$ and $A_2$ [see Fig. \ref{fig:1D_chain}(b)]
after tracing out $B$ is zero.
We can understand the vanishing of the entanglement negativity from the following picture.
All the single-particle states are localized on the dimerized bonds that 
no single-particle states can spread out on both $A_1$ and $A_2$.
Hence there is no entanglement between $A_1$ and $A_2$.

\subsubsection{Integer Quantum Hall state (IQHS)}
\label{Sec: IQHS}
\begin{figure}[htbp]
\centering
   \includegraphics[height=1.1 cm] {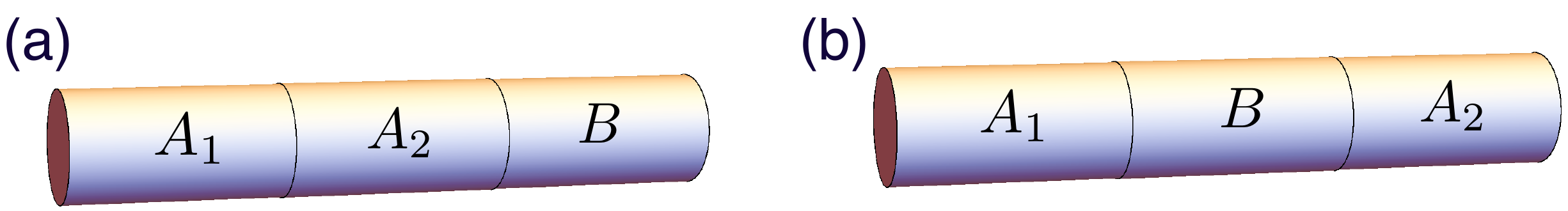}
\caption{The tripartite configuration of a integer quantum Hall state on a cylinder. 
The system is divided into $A_1$, $A_2$, and $B$.
(a) Adjacent intervals for $A_1$ and $A_2$ regions.
(b) Disjointed intervals for $A_1$ and $A_2$ regions.
}
\label{fig:IQHE_1}
\end{figure}
We consider the IQHS on a cylinder with the periodic boundary condition along $y$ direction.
The entanglement entropy/spectrum analysis on IQHS on a cylinder has been studied in Refs. [\onlinecite{Ivan, Chang}].
The single-particle state in the lowest Landau level
with the Landau gauge is
\begin{align}
\psi_{k_y}(x,y)=\frac{1}{\pi^{1/4} L_y^{1/2}}e^{ik_y y} e^{-(x-l_B^2 k_y)^2/(2l_B^2)},
\label{eq: iqhs}
\end{align}
where 
\begin{align}
k_y = - \frac{L_x}{2}+ n \frac{ \pi l_B^2 }{L_y}, \quad  1 \leqslant n \leqslant \frac{L_x L_y}{ \pi l_B^2}.
\end{align}
The particles are restricted in the region $|x|\leqslant L_x/2$ and the number of particles is $\frac{L_xL_y}{\pi l_B^2}$ (we set
the magnetic length $l_B=1$).
The system is divided
into $A_1$, $A_2$, and $B$.
For the case of two adjacent intervals $A_1$ and $A_2$.
We have $A_1 \in [-\infty, -L_1]$,
$A_2 \in [-L_1, L_2]$, and $B \in [ L_2, \infty]$ along $x$ axis [Fig. \ref{fig:IQHE_1}(a)].
From Refs. [\onlinecite{Ivan, Chang}],
the corresponding probabilities for a particle at regions $A_1$, $A_2$ and $B$ are

\begin{align}
&P_{A_1}=\frac{1}{\sqrt{\pi}} \int_{-\infty}^{-L_1} dx e^{-(x-k_y)^2}    = \frac{1}{2}(1-\mathrm{erf}(k_y+L_1)),      \notag\\        
&P_{A_2}=1-P_{A_1}-P_{B}=\frac{1}{2}(\mathrm{erf}(k_y+L_1)-\mathrm{erf}(k_y-L_2)), \notag\\       
&P_{B}=\frac{1}{\sqrt{\pi}} \int_{L_2}^{\infty} dx e^{-(x-k_y)^2}  = \frac{1}{2}(1+\mathrm{erf}(k_y-L_2)). 
\label{erf1}
\end{align}

For the case of two disjointed intervals $A_1$ and $A_2$.
We have $A_1 \in [-\infty, -L_1]$,
$B \in [-L_1, L_2]$, and $A_2 \in [ L_2, \infty]$ along $x$ axis [Fig. \ref{fig:IQHE_1}(b)].
The corresponding probabilities for a particle at regions $A_1$, $B$, and $A_2$ are
\begin{align}
&P_{A_1}= \frac{1}{2}(1-\mathrm{erf}(k_y+L_1)),      \notag\\        
&P_{B}=\frac{1}{2}(\mathrm{erf}(k_y+L_1)-\mathrm{erf}(k_y-L_2)), \notag\\       
&P_{A_2}= \frac{1}{2}(1+\mathrm{erf}(k_y-L_2)). 
\label{erf2}
\end{align}

For simplicity, we set $L_1=L_2 \ll L_x$.
From Eq. (\ref{Eq: EN for mixed state}), the entanglement negativity between $A_1$ and $A_2$
after tracing out $B$ is
\begin{align}
\mathcal{E} =& \sum_{k_{y}}\ln (P_{A_1}(k_y) + P_{A_2}(k_y) \notag\\
&+ \sqrt{(P_{B}(k_y))^2+4P_{A_1}(k_y) P_{A_2}(k_y)}).
\end{align}
We can also compute 
the slope of the entanglement negativity by taking the continuous limit 
\begin{align}
\frac{\mathcal{E}(L_y)}{L_y} 
=&\frac{1}{ \pi}\int_{-L_x/2}^{L_x/2} \ln (P_{A_1}(k_y) + P_{A_2}(k_y)  \notag\\
&+ \sqrt{(P_{B}(k_y))^2+4P_{A_1}(k_y) P_{A_2}(k_y)})dk_y.  
\label{Eq: slope}
\end{align}

\paragraph{Adjacent intervals}
\begin{figure}[htbp]
\centering
\includegraphics[height=3.7 cm] {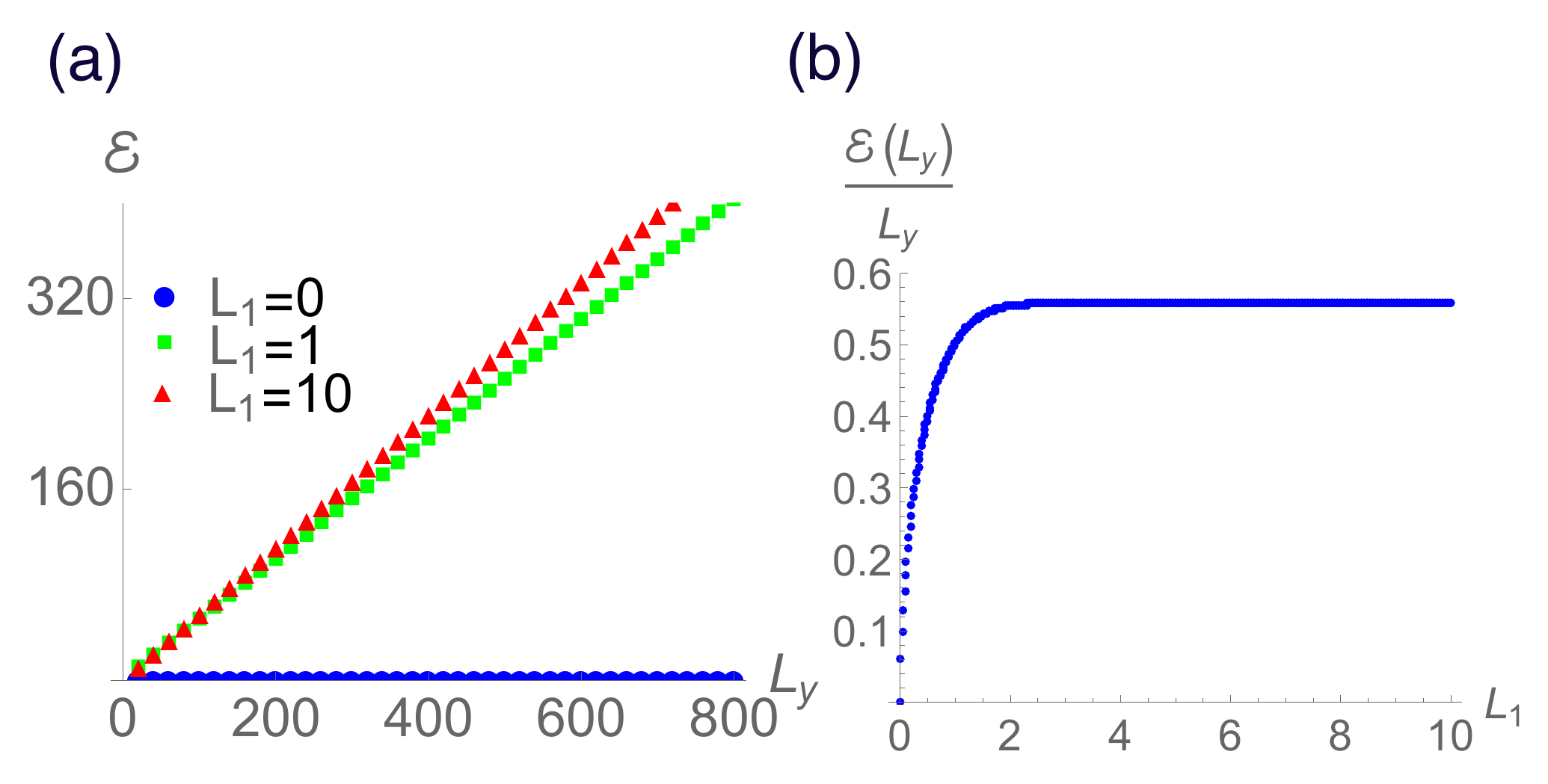}
\caption{
The entanglement negativity $\mathcal{E}$ for two adjacent intervals ($A_1$ and $A_2$) in the integer quantum Hall state on a cylinder.
The length of subsystem $A_2$ is $L_1$ and the compactified length 
on the cylinder is $L_y$.
(a) $\mathcal{E}(L_y)$ for $L_1=0$ (Blue),  $L_1=1$ (Green), and $L_1=10$ (Red).
(b) $\frac{\mathcal{E}(L_y)}{L_y}$ as a function of $L_1$. We set $L_x= 100 \pi$ that particles are restricted in the region $x  \leqslant |L_x/2|$.
}
\label{fig:Jointed_IQHE}
\end{figure}
The entanglement negativity is linearly proportional to $L_y$ [Fig. \ref{fig:Jointed_IQHE} (a)], i.e., an area law behavior.
For $L_1=0$, the dimension of Hilbert space of region $A_2$ is zero that leads to the vanishing of the entanglement negativity.
In the limit that $L_1 \ll L_x$, the entanglement negativity increases monotonically as $L_1$ increasing. 
Notice that when $L_1 > L_x$, all the particles are restricted in the subsystem $A_2$.
Hence the entanglement negativity goes to zero. The slope of the entanglement
computed from Eq. (\ref{Eq: slope}) is shown in Fig. \ref{fig:Jointed_IQHE} (b).
The plateau of $\frac{\mathcal{E}(L_y)}{L_y}$ is around  $0.56$,
which indicates the system approaches to the case of a pure state when $L_1 \gg 1$. 
We demonstrate the case of a pure state in the IQHS in the Appendix. \ref{App_IQHE_pure}.

\begin{figure}[htbp]
\centering
   \includegraphics[height=4 cm] {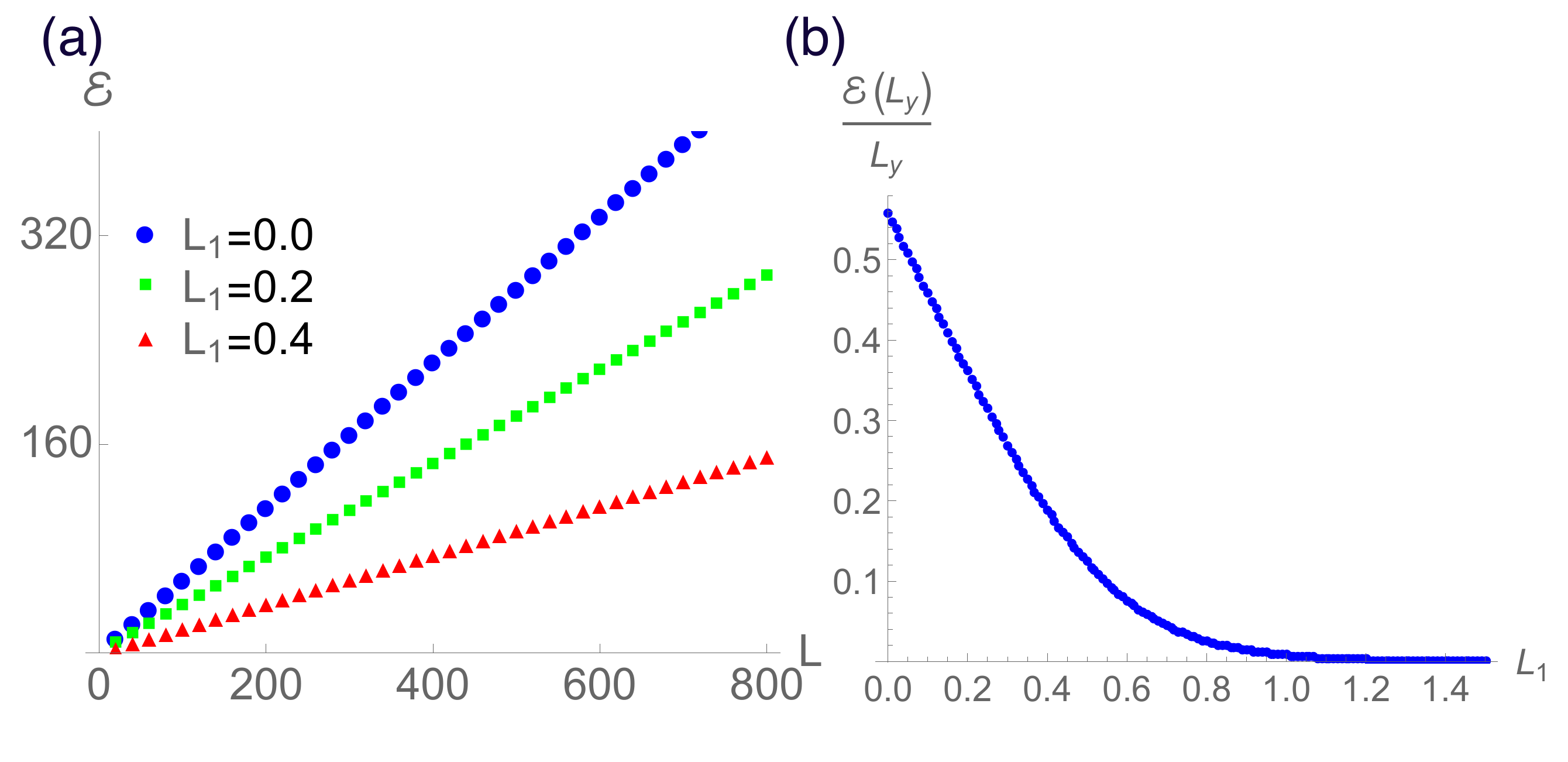}
\caption{ The entanglement negativity $\mathcal{E}$ for two disjointed intervals ($A_1$ and $A_2$) in the integer quantum Hall state on a cylinder.
The length of subsystem $B$ and the compactified length on
on the cylinder is $L_y$.
(a) $\mathcal{E}(L_y)$ for $L_1=0$ (blue),  $L_1=0.2$ (green), and $L_1=0.4$ (red).
(b) $\frac{\mathcal{E}(L_y)}{L_y}$ as a function of $L_1$. We set $L_x = 100 \pi$ that particles are restricted in the region $x  \leqslant |L_x/2|$.
}
\label{fig:Disjointed_IQHE}
\end{figure}

\paragraph{Disjointed intervals}
In Fig. \ref{fig:Disjointed_IQHE} (a), the entanglement negativity also shows an area law behavior as the case of adjacent intervals.
For $L_1=0$, the dimension of Hilbert space of region $B$ is zero. 
Hence the entanglement negativity is exact the same as the entanglement negativity for a pure state. 
As $L_1$ is increasing, the entanglement negativity decreases. This indicates that $A_1$ and $A_2$ become less entangled when
the distance between these two intervals becomes larger.
The slope of the entanglement
computed from Eq. (\ref{Eq: slope}) is shown in Fig. \ref{fig:Disjointed_IQHE} (b).
When $L_1=0$, the system can be described by a pure state and $\frac{\mathcal{E}(L_y)}{L_y} \sim 0.56 \cdots$.
Since the decay length in the system is proportional to $l_B$,
$A_1$ and $A_2$ become less entangled when the distance between two disjointed intervals $L_1$ is greater then $l_B$.
We find that $\frac{\mathcal{E}(L_y)}{L_y}$ monotonically decreases as $L_1$ is increasing.

\subsubsection{Discussion} 
For the IQHS, as the distance $L_1$ is larger than the magnetic length $l_B=1$, the entanglement negativity vanishes dramatically. 
It may be straightforwardly understood in the following. As studied in Ref. [\onlinecite{Qi2011}], 
for a (2+1) dimensional topological quantum state which possesses chiral edge states, such as the IQHS we discussed here, 
the reduced density matrix of a spatial region can be expressed as the density matrix of the corresponding edge states. 
In other words, the entanglement is mainly contributed by the boundary states which are localized at the interface. 
Therefore, for the configuration in Fig. \ref{fig:IQHE_1}, the reduced density matrix $\rho_{A_1\cup A_2}$ can be expressed as
\begin{equation}
\rho_{A_1\cup A_2}=\rho_{\partial A_1}\otimes \rho_{\partial A_2},
\end{equation}
where $\rho_{\partial A_i}$ represents the density matrix corresponding to the boundary states located at the interface $\partial A_i$ between region $A_i$ and region $B$. In this case, one has
\begin{equation}
\rho_{A_1\cup A_2}^{T_2}=\rho_{\partial A_1}\otimes \rho_{\partial A_2}^{T},
\end{equation}
which leads to a vanishing entanglement negativity. Our conclusion also applies to a general topological ordered states such as the fractional quantum hall effect. 
The topological flux threading through the cylinder in Fig. \ref{fig:IQHE_1} can be in a superposition
of different topological sectors, and the reduced density matrix has the expression
\begin{equation}
\rho_{A_1\cup A_2}=\sum_a |\psi_a|^2 \rho_{\partial A_1}^a\otimes \rho_{\partial A_2}^a,
\end{equation}
where $|\psi_a|^2$ is the probability corresponding to the topological sector $a$. Again,  the state is explicitly separable, 
and therefore the entanglement negativity is zero. 
Note that the above conclusion was also obtained in Ref. [\onlinecite{Lee}] based on the solvable toric code mode, 
but here we give an argument which holds for a general topological ordered state in (2+1)D. 
A more strict proof based on the boundary state picture can be found in Ref. [\onlinecite{Wen2015}].
What ia interesting, most recently, the mutual information between $A_1$ and $A_2$ in  Fig. \ref{fig:IQHE_1} for a (2+1) dimensional topological
ordered state was studied \cite{Qi2015}. It was found that 
\begin{equation}\label{IA1A2}
I_{A_1A_2}=-\sum_a|\psi_a|^2\mathrm{ln}|\psi_a|^2,
\end{equation}
which is nonzero. One may understand the difference between the entanglement negativity and the mutual information 
in the following way. As pointed out
in Ref. [\onlinecite{Qi2015}], the nonzero contribution in Eq. (\ref{IA1A2}) comes from the ``classical" correlation between extended objects.
It is known that the mutual information contains both classical and quantum correlations, while the entanglement 
negativity measures the completely quantum part of correlations\cite{QiEmail, Castelnovo}. Therefore, one may have a vanishing entanglement negativity while the mutual information is finite.

\subsection{
The ground state in a tripartite system cannot be factored into triples of modes}
\label{Sec: EE_mix_nondiag}
For the case where the overlap matrices cannot be simultaneous diagonalized in a tripartite system,
the ground state cannot be factorized.
Although the ground state cannot be factorized, we can express the ground state
in terms of the fermion operators in each subsystem.
The procedure of constructing the ground state is the following.
We start with the bipartite ground state
\begin{align}
|\Psi  \rangle=&  \prod_{i=1}^{N} (\sqrt{P_i} d^\dagger_{Ai}+ \sqrt{1-P_i}d^\dagger_{Bi}) |0 \rangle.
\end{align} 
The orthonormal modes at $A$ are the set $\{ |A_i \rangle; d_{Ai}^\dagger | 0 \rangle = |A_i \rangle\}$.
Next, we use these orthonormal modes in subsystem $A$ to construct
the overlap matrices $M_{A_1}$ and $M_{A_2}$,
\begin{align}
[M_{A_i}]_{\alpha, \beta} =\langle \mathcal{P}_{A_{i}} A_\alpha,  \mathcal{P}_{A_{i}} A_\beta \rangle,\quad i=1,2,
\end{align}
where $\mathcal{P}_{A_i}$ is the orthogonal projection operator on Hilbert space $A_{i}$.
In other words, the tripartite system is built from partitioning the system into $A$ and $B$ parts first,
and partitioning the subsystem $A$ into $A_1$ and $A_2$ parts.
Since $\{|A_i \rangle \}$ is the orthonormal set, we have $M_{A_1} +M_{A_2} = \mathbb{I}$.
We can simultaneously diagonalize $M_{A_1}$ and $M_{A_2}$ by a unitary matrix $V$,
such that $VM_{A_1} V^\dagger = \rm{diag}(Q_k)$ and $VM_{A_2} V^\dagger = \rm{diag}(1-Q_k)$.
The orthonormal modes in $A_1$ and $A_2$ are
\begin{align}
|A_{1k} \rangle=\frac{V^\dagger _{k i} \mathcal{P}_{A_1} | A_i \rangle}{\sqrt{Q_k}},  \quad
|A_{2k} \rangle=\frac{V^\dagger _{k i} \mathcal{P}_{A_2} | A_i \rangle}{\sqrt{1-Q_k}}.
\end{align}

Now the ground state can be expressed as
\begin{align}
|\Psi  \rangle=&  \prod_{i=1}^{N} (\sqrt{P_i} d^\dagger_{Ai}+ \sqrt{1-P_i}d^\dagger_{Bi}) |0 \rangle \notag\\
=& \prod_{i=1}^{N} (\sqrt{P_i } \sum_{k} V_{i k} (\sqrt{Q_k}d^\dagger_{A_1 k}+\sqrt{1-Q_k}d^\dagger_{A_2 k} ) \notag\\
&+ \sqrt{1-P_i}d^\dagger_{Bi}) |0 \rangle,
\label{Eq:gwf2}
\end{align}
where $d_{A_i k}^\dagger |0 \rangle = |A_{ik} \rangle $ with $i=1,2$.
The second line in the above equation comes from 
\begin{align}
d^{\dagger}_{Ai}|0\rangle &= |A_i \rangle = \mathcal{P}_{A_1} |A_i \rangle +   \mathcal{P}_{A_2} |A_i \rangle   \notag\\
&=\sum_{k} V_{ik} (\sqrt{Q_k}d^\dagger_{A_1 k}+\sqrt{1-Q_k}d^\dagger_{A_2 k} )|0 \rangle.
\end{align}

In terms of occupation basis, the ground state can be expressed as
\begin{align}
|\Psi  \rangle= \sum_{n=0}^{N} C_{n}| n_{A}, (N-n)_B \rangle,
\end{align}
where $n_A$ indicates that there are $n$ electrons in subsystem $A= A_1 \cup A_2$
and $(N-n)_B$ indicates that there are $N-n$ electrons in subsystem $B$.
We can construct the partially transposed reduced density matrix from the occupational basis [see detail derivations in App. \ref{App_Nontri}].
The final expression of $\rho_A^{T_{A_2}} $ is 
\begin{align}
\rho_A^{T_{A_2}} = \sum_{n_i=11, 00,10,01} C_{n_1, n_2, \cdots ,n_N} \bigotimes_{i=1}^{N} [\rho_{n_i}]^{T_{A_2}}_i,
\label{Eq: general}
\end{align}
where $\rho_{n_i}$ are four elemental matrices (see Eq. (\ref{Eq: 4M}) in App. \ref{App_Nontri}) as the building blocks of $\rho_A^{T_{A_2}}$.
Thus, the entanglement negativity can be obtained by taking the trace norm,
 $\mathcal{E} = \ln \mathrm{Tr} |\rho_A^{T_{A_2}}|$.

\subsubsection{Free fermions on a 1D ring}
\begin{figure}[htbp]
\centering
   \includegraphics[height=3.2 cm] {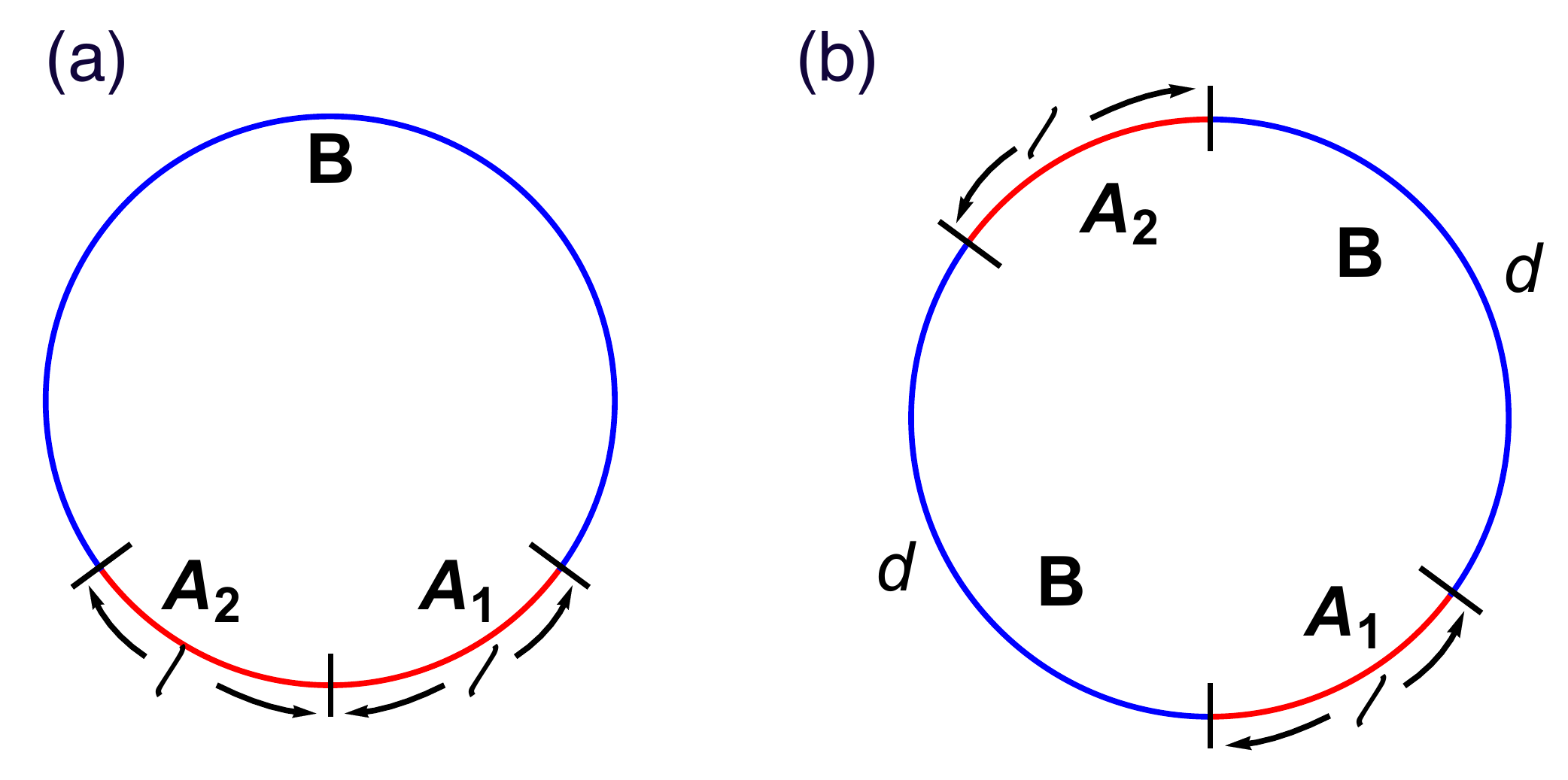}
\caption{ Two tripartite configurations of 1D rings: 
(a) Two adjacent intervals $A_1$ and $A_2$ with equal length $l$.
(b) Two disjointed intervals $A_1$ and $A_2$ with equal length $l$ and the length between these
two intervals is $d$.  
}
\label{fig: EN_rings}
\end{figure}

We consider free fermions on a periodic one-dimensional chain.
The normalized single-particle eigenstate is 
\begin{align}
\phi_{n} = \frac{1}{\sqrt{L}} e^{i \frac{2 n \pi}{L} x}, \quad n \in \mathbb{Z},
\end{align}
with energy $E_n = (\frac{2 \pi n}{L})^2$.
The overlap matrix can be constructed from these eigenstates from Eq. (\ref{Eq: overlap_matrix})
and the partially transposed reduced density matrix can be obtained from Eq. (\ref{Eq: general}).

Due to the limitation of numerical calculation, we consider few fermions (up to five) on a long chain ($L > 50$).
For simplicity, two intervals $A_1$ and $A_2$ have equal length $l$ and
the total length of the ring is $L$.

\paragraph{Adjacent intervals}

\begin{figure}[htbp]
\centering
   \includegraphics[height=4. cm] {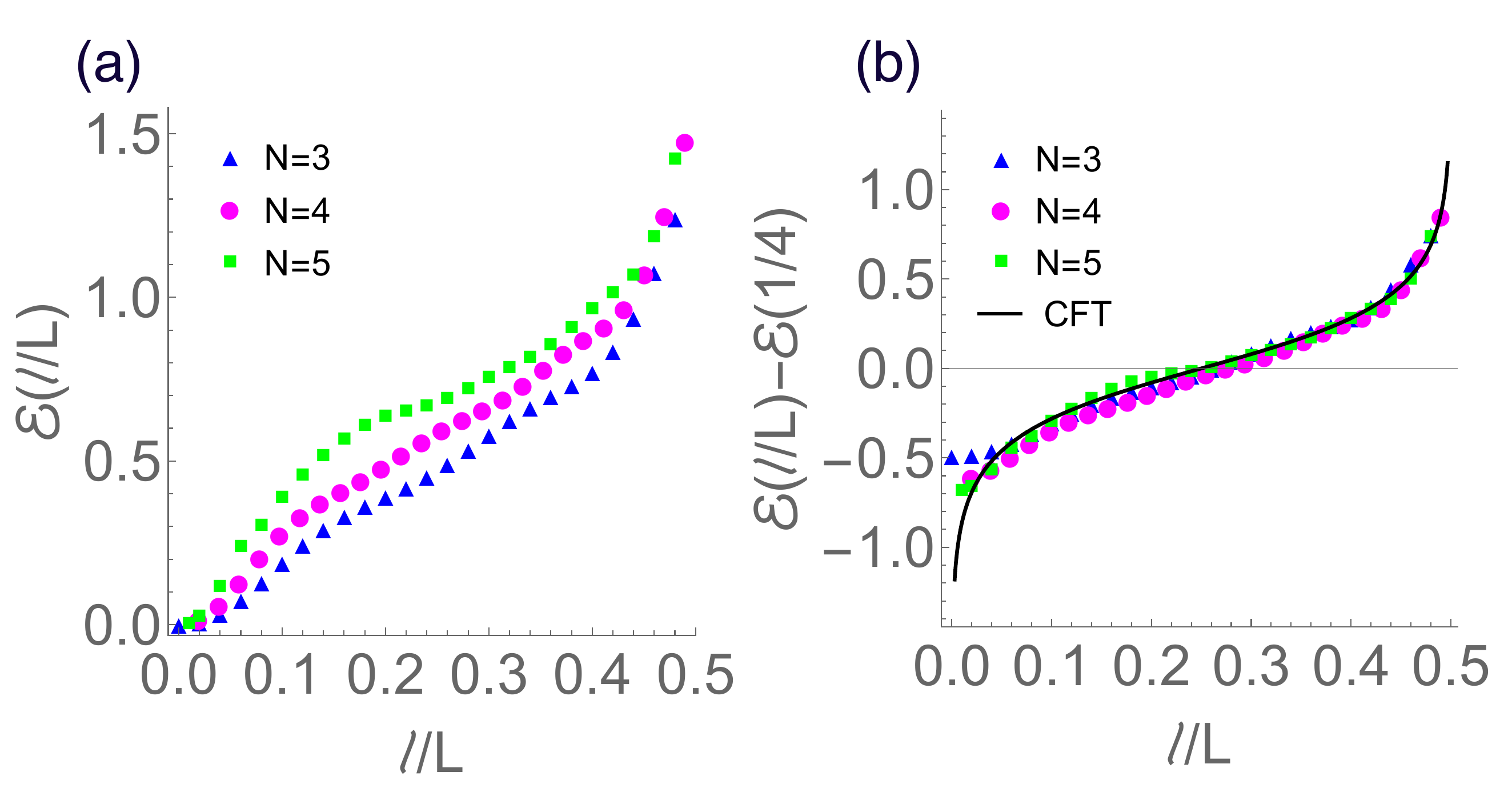}
\caption{(a) The entanglement negativity $\mathcal{E}(l/L)$ as a function of $l/L$ 
for two adjacent intervals in a homogeneous one-dimensional ring with $N$ particles.
Two intervals are equal length $l$ and the total length of the ring is $L$.
 (b) $\mathcal{E}(l/L)-\mathcal{E}(1/4)$ as a function of $l/L$. 
Solid grey line is the scaling function from CFT calculation, $\mathcal{E}= \frac{1}{4} \mathrm{ln[tan}(\pi \frac{l}{L}))]$. 
}
\label{fig: EN_CFT}
\end{figure}
The configuration of adjacent intervals is shown in Fig. \ref{fig: EN_rings}(a).
For fixed number of particles, the entanglement negativity of adjacent intervals depends on the single parameter $l/L$.
The entanglement negativity increases monotonically as the function of $l/L$  [Fig. \ref{fig: EN_CFT} (a)].
In the CFT calculation\cite{Calabrese,Calabrese2}, $\mathcal{E}= \frac{1}{4} \mathrm{ln[tan}(\pi \frac{l}{L}))]+ \mathrm{const.}$.
In order to compare with the CFT calculation and get rid of the constant part in the entanglement negativity, 
we plot $\mathcal{E}(l/L)-\mathcal{E}(1/4)$ for $N=3,4,5$.
in Fig. \ref{fig: EN_CFT} (b). The entanglement negativity after the subtraction of $\mathcal{E}(1/4)$ agrees with the CFT prediction.

\begin{figure}[htbp]
\centering
   \includegraphics[height=3.2 cm] {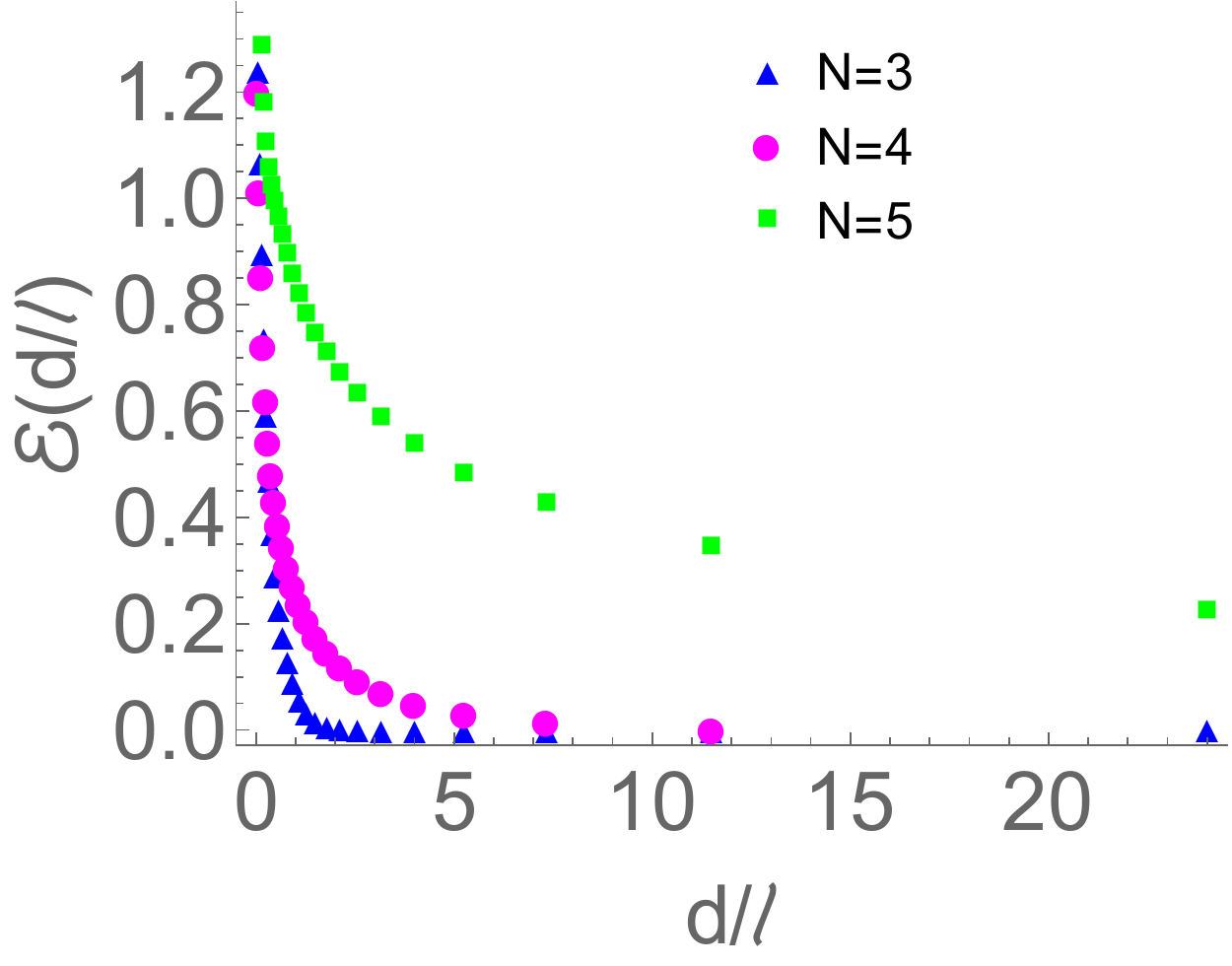}
\caption{ The entanglement negativity $\mathcal{E}(d/l)$ as a function of $d/l$ for two disjointed intervals in a homogeneous one-dimensional ring with $N$ particles.
Two disjointed intervals have equal length $l$ and the distance between two intervals is $d$.
}
\label{fig: EN_disjointed}
\end{figure}

 \paragraph{Disjointed intervals}
 The entanglement negativity of disjointed intervals for harmonic chains has been studied in Refs. [\onlinecite{Wichterich2,Marcovitch,Calabrese}].
 Here, we consider a symmetric configuration of disjointed intervals as shown in Fig. \ref{fig: EN_rings}(b).
 For fixed number of particles, the entanglement negativity of disjointed intervals depends on the single parameter $d/l$, where $d$
 is the distance between two intervals.
 The entanglement negativity decreases monotonically as the function of $d/l$ [Fig. \ref{fig: EN_disjointed}].

\section{Discussion and conclusion}
\label{Sec: conclusion}
In this paper, we systematically derive 
the entanglement negativity in free-fermion systems by using the overlap matrices.
For a pure state, the entanglement negativity is obtained from a bipartite system and 
the ground state can always be factored into pairs of modes.
For a mixed state, the factorability of the ground state in a tripartite system does not always hold.
In the case that the ground state can still be factored into triples of modes, 
the entanglement negativity has a simple form. We show two examples in this case: the dimerized SSH model
and the IQHS. The entanglement negativity for the former is $\ln 2$ when two
intervals are adjacent. The entanglement negativity for the latter shows an area law behavior.
For both examples, we observe that the entanglement negativity 
vanishes when the separation between two intervals is greater 
than the correlation length of the systems.
We argue the vanish of the entanglement negativity holds for a general topological ordered
state in $(2+1)$D. A proof based on the edge theory approach as well as the bulk theory calculation
is given in Refs. [\onlinecite{Wen2015}] and Ref. [\onlinecite{EN_CS}], respectively.
On the other hand, if the ground state {\it cannot} be factored into triples of modes, we 
show that the partially transposed reduced density matrix can be expressed in terms of the Kronecker product 
of matrices. In this case, we compute the entanglement negativity for free fermions on a one-dimensional ring.
We find that the entanglement negativity of two adjacent intervals agrees with the CFT prediction.


We would like to end the conclusion by mentioning a few of future problems.

$\bullet$ It is interesting to study the non-equilibrium dynamics of entanglement negativity for a free-fermion system. 
In particular, as discussed in Ref. [\onlinecite{Wen}], for the time evolution of entanglement negativity $\mathcal{E}(t)$ 
after a local quench, the result is not universal for some region $t$, i.e., $\mathcal{E}(t)$ may depend on the concrete 
lattice models. It is interesting to check the behavior of $\mathcal{E}(t)$ for a critical free fermion chain, and compare it with 
the results of other lattice models \cite{Eisler}. 

$\bullet$ In our current work, we mainly focus on the free-fermion systems at zero temperature. 
It is desirable to generalize our method to the finite temperature case, and compare with the CFT results\cite{Calabrese4}. 

$\bullet$ One can easily apply our methods to free-fermion systems with disorder.
Studying the quantum phase transitions of disordered systems by using
the entanglement negativity is needed for a future investigation.

$\bullet$ It is interesting to extend our method to the study of interacting fermionic systems, such as fractional quantum hall systems. 
Although there is no entanglement negativity for two disjoint intervals, it is still interesting to study the entanglement
negativity for two adjacent intervals, where we may have a nonvanishing contribution from the long-range entanglement.

\section{Acknowledgement}
The authors 
thank Shinsei Ryu for his encouragement and discussions
during this work.
PYC is supported by the Rutgers Center for Materials Theory group postdoc grant.


 

\appendix

\section{Entanglement negativity for pure states}
\label{App: EE_pure}

Let us consider a bipartite system 
that the ground state can be {\it{Schmidt}} decomposed as
\begin{align}
|\Psi  \rangle = \sum_{i}C_{i} |\psi_{Ai} \psi_{Bi} \rangle,
\end{align}
where $C_i$ is the coefficient that in between zero and one and the system is bipartite into A and B parts.
The density matrix and the reduced density matrix are defined as
\begin{align}
&\rho =|\Psi  \rangle \langle \Psi | = \sum_{i,j} C_{i} C_{j} |\psi_{Ai} \psi_{Bi}  \rangle \langle \psi_{Aj}\psi_{Bj} |,  \notag \\
&\rho_A=\mathrm{Tr}_{B} \rho = \sum_{i} C_i^2 |\psi_{Ai}  \rangle \langle \psi_{Ai} |.
\end{align}

One can define a partial transpose of the density matrix as
\begin{align}
\rho^{T_B}= \sum_{i,j} C_{i} C_{j} |\psi_{Ai} \psi_{Bj}  \rangle \langle \psi_{Aj}\psi_{Bi} |.
\end{align}
The trace norm of $\rho^{T_B}$ is defined by
\begin{align}
\mathrm{Tr}|\rho^{T_B}| :=\sum_{i} |\Lambda_i|,
\end{align}
where $\Lambda_i$ is the eigenvalue of $\rho^{T_B}$.

The (logarithmic) entanglement negativity is defined as
\begin{align}
\mathcal{E} :=\mathrm{ln \, Tr}  |\rho^{T_B}|.
\end{align}





\subsection{Su-Schrieffer-Heeger (SSH) model}

\begin{figure}[htbp]
\centering
   \includegraphics[height=3.2 cm] {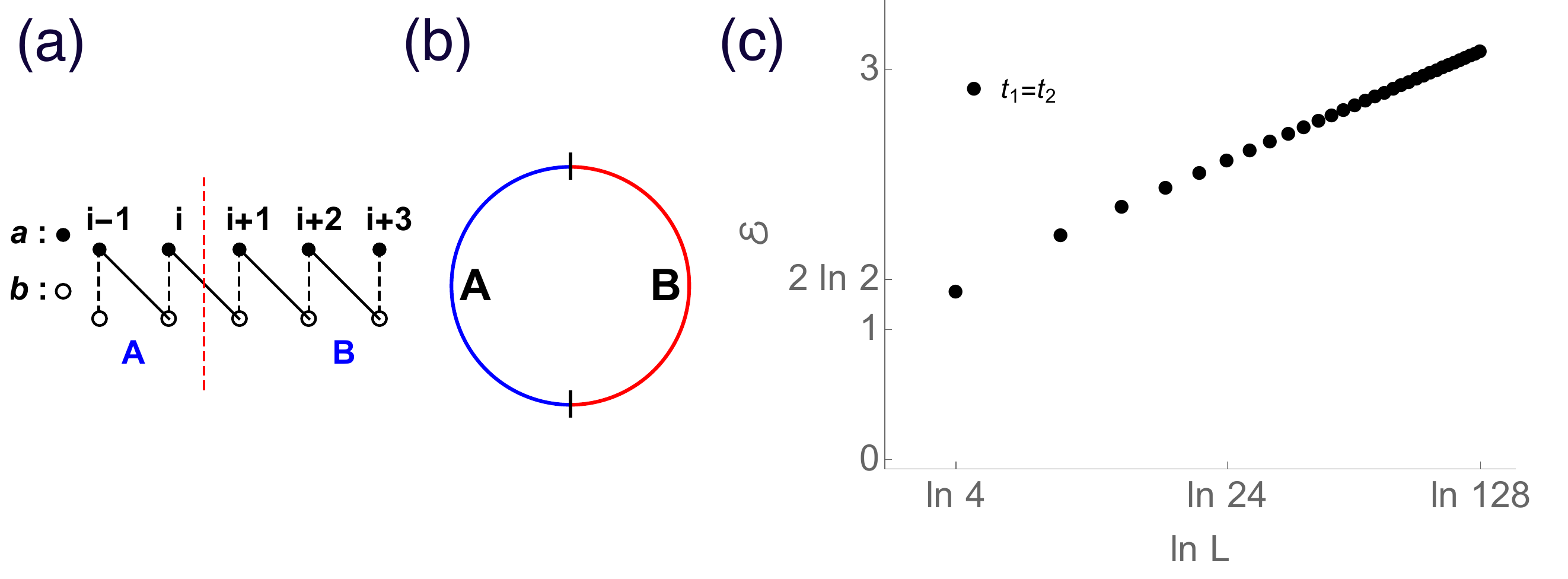}
\caption{(a) Su-Schrieffer-Heeger model on a one-dimensional ring. The hopping amplitude of dashed(solid) lines is $t_1(t_2)$. 
A unit cell contains two orbitals that are pictured by $\bullet$ and $\circ$. The system is divided 
into $A$ and $B$ parts by the red dashed line.
(b) The bipartite configuration of a ring with the lengths of $A$ and $B$ being equal.  
(c) The entanglement negativity $\mathcal{E}$ for an equally bipartite Su-Schrieffer-Heeger model with $t_1=t_2$ as a function of $\mathrm{ln}L$, where $L$ is the length
of the ring. 
}
\label{fig: 1Dring}
\end{figure}

Let us consider an one-dimensional tight binding Hamiltonian (SSH model) under periodic boundary condition
\begin{align}
H=\sum_i t_1 c^\dagger_{a_i} c_{b_i}+ t_2  c^\dagger_{a_{i+1}} c_{b_{i}} + \mathrm{h.c.} \notag\\
\label{eq: ssh}
\end{align}
where $t_1,t_2>1$ and $c_{a(b)_i}$ is the fermion operator. 
The configuration of a bipartite ring is shown in Fig. \ref{fig: 1Dring}.
We consider the lengths of $A$ and $B$ are equal.
The overlap matrix $M_A$ has two eigenstates with $0.5$ eigenvalue when $t_2>t_1$.
On the other hand, there is no eigenstates of $M_A$ with $0.5$ eigenvalue when $t_1>t_2$.
For $t_1=0$, the set of the eigenvalues of $M_A$ is $\{P_i\}=\{0,0,\cdots,0.5,0,5,1,1,\cdots,1 \}$, that leads to $\mathcal{E}= 2 \mathrm{ln} 2$.
For $t_2=0$, the set of the entanglement spectrum is $\{P_i\}=\{0,0,\cdots,0,1,\cdots,1 \}$.
This implies $\mathcal{E}=0$.
At critical point, $t_1=t_2$, the scaling function of the entanglement negativity is proportional to $\mathrm{ln} L$ 
with $L$ being the length of the system. 
The set of points in Fig. \ref{fig: 1Dring}(c) fits $0.71+0.50 x$.
From the CFT prediction\cite{Calabrese}, the entanglement negativity is
$\mathcal{E}=\frac{c}{2} \ln \frac{L}{\pi } \sin \frac{\pi l}{L}+ \mathrm{const.}$.
In our consideration, $l/L=1/2$, that leads to $\mathcal{E}=\frac{c}{2} \ln L + \mathrm{const.}$. 
One can read off the central charge $c \sim 1$.

\subsection{Integer Quantum Hall state (IQHS)}
\label{App_IQHE_pure}
We consider the integer quantum Hall state (IQHS) on a cylinder.
We choose the periodic boundary condition along $y$ direction.
Hence $k_y$ is a good quantum number. 
The single-particle state in the lowest Landau level
with the Landau gauge is
\begin{align}
\psi_{k_y}(x,y)=\frac{1}{\pi^{1/4} L_y^{1/2}}e^{ik_y y} e^{-(x-l_B^2 k_y)^2/(2l_B^2)},
\label{eq: iqhs}
\end{align}
where 
\begin{align}
k_y = - \frac{L_x}{2}+ n \frac{ \pi l_B^2 }{L_y}, \quad  1 \leqslant n \leqslant \frac{L_x L_y}{ \pi l_B^2}.
\end{align}
Here the number of particles $ \frac{L_x L_y}{ \pi l_B^2}$ is obtained by restricting the particles in
the region $|x| \leqslant L_x/2$. Without loss of generality, we set the magnetic length $l_B=1$.
The system is divided into $A$ and $B$ at $x=0$, where $A \in [-\infty, 0]$ and $B \in [0, \infty]$ along $x$ axis.
From Refs. [\onlinecite{Ivan, Chang}], 
the corresponding probabilities for a particle at region $A$ and $B$ are

\begin{align}
&P_{A}=\frac{1}{\sqrt{\pi}} \int_{-\infty}^{0} dx e^{-(x-k_y)^2}    = \frac{1}{2}(1-\mathrm{erf}(k_y)),   \notag\\        
&P_{B}=\frac{1}{\sqrt{\pi}} \int_{0}^{\infty} dx e^{-(x-k_y)^2}  = \frac{1}{2}(1+\mathrm{erf}(k_y)). 
\label{Eq: erf}
\end{align}

From Eq. (\ref{Eq: pure}), the entanglement negativity is
\begin{align}
\mathcal{E}=\sum_{k_y} \mathrm{ln}( 1+ \sqrt{1- \mathrm{erf}(k_y)^2}).
\end{align}

Numerically we observe that the entanglement negativity is linearly proportional to the length $L_y$.
This indicates that the entanglement negativity satisfies an area law.
The slope of the entanglement negativity can be obtained by taking the continuous limit
\begin{align}
\frac{\mathcal{E}(L_y)}{L_y} &=\lim_{L_x \to \infty} \frac{1}{ \pi}\int_{-L_x/2}^{L_x/2} \ln [1+ \sqrt{1- \mathrm{erf}^2(k_y)}]dk_y   \notag\\
&\simeq 0.56\cdots.
\end{align}

\section{Detail expression of the partially transposed reduced density matrix for the non-triply factored case}
\label{App_Nontri}
In the main text, we can write down the ground state
in a tripartite systems as,
\begin{align}
|\Psi  \rangle=&  \prod_{i=1}^{N} (\sqrt{P_i} d^\dagger_{Ai}+ \sqrt{1-P_i}d^\dagger_{Bi}) |0 \rangle \notag\\
=& \prod_{i=1}^{N} (\sqrt{P_i } \sum_{k} V_{i k} (\sqrt{Q_k}d^\dagger_{A_1 k}+\sqrt{1-Q_k}d^\dagger_{A_2 k} ) \notag\\
&+ \sqrt{1-P_i}d^\dagger_{Bi}) |0 \rangle,
\label{AppEq:gwf2}
\end{align}
where $V_{ik}$ is a unitary matrix that can simultaneously diagonalize the overlap matrices $M_{A_1}$ and $M_{A_2}$
in region $A= A_1 \cup A_2$.
In terms of occupation basis, the ground state can be expressed as
\begin{align}
|\Psi  \rangle= \sum_{n=0}^{N} C_{n}| n_{A}, (N-n)_B \rangle,
\end{align}
where $n_A$ indicates that there are $n$ electrons in subsystem $A= A_1 \cup A_2$
and $(N-n)_B$ indicates that there are $N-n$ electrons in subsystem $B$.
We have

\begin{widetext}

\begin{align}
C_{n}| n_{A}, (N-n)_B \rangle =& \sum_{i_1>i_2>\cdots>i_n}
\sqrt{P_{i_1}}\sqrt{P_{i_2}} \cdots \sqrt{P_{i_n}}  \sum_{k_1>k_2>\cdots>k_n}  \mathrm{Det}[V^{\mathrm{minor}}_{\{i\},\{k\}}] \notag\\
&\times
(\sqrt{Q_{k_1}}d^{\dagger}_{A_1k_1} + \sqrt{1-Q_{k_1}}d^{\dagger}_{A_2 k_1})
(\sqrt{Q_{k_2}}d^{\dagger}_{A_1k_2} + \sqrt{1-Q_{k_2}}d^{\dagger}_{A_2 k_2}) \cdots  \notag\\
&\times
(\sqrt{Q_{k_n}}d^{\dagger}_{A_1k_n} + \sqrt{1-Q_{k_n}}d^{\dagger}_{A_2 k_n})
\prod_{\alpha \neq i_1, i_2,\cdots,i_n}\sqrt{1-P_{\alpha}} d^{\dagger}_{B\alpha}
 |0 \rangle,  
\end{align}
where $\mathrm{Det}[V^{\mathrm{minor}}_{\{i\},\{k\}}]$ is the determinant of the unitary matrix $V$
by picking $i_1,i_2,\cdots,i_n$-th rows and  $k_1,k_2,\cdots,k_n$-th columns of $V$.

The reduced density matrix after tracing $B$ part is
$\rho_A = \sum_{n=0}^{N} D_{n} |n_A  \rangle \langle n_A |$,
where
\begin{align}
D_{n} |n_A  \rangle \langle n_A | =& \sum_{i_1>i_2>\cdots>i_n} 
P_{i_1}P_{i_2} \cdots P_{i_n}\prod_{\alpha \neq i_1, i_2,\cdots,i_n}(1- P_{\alpha}) 
 \sum_{k_1>k_2>\cdots>k_n}  \sum_{k'_1>k'_2>\cdots>k'_n}  \mathrm{Det}[V^{\mathrm{minor}}_{\{i\},\{k\}}]  
  \mathrm{Det}[V^{\mathrm{minor}}_{\{i\},\{k'\}}]  \notag\\
& \times
(\sqrt{Q_{k_1}}d^{\dagger}_{A_1k_1} + \sqrt{1-Q_{k_1}}d^{\dagger}_{A_2 k_1})
\cdots  (\sqrt{Q_{k_n}}d^{\dagger}_{A_1k_n} + \sqrt{1-Q_{k_n}}d^{\dagger}_{A_2 k_n})  |0 \rangle  \notag\\
&\times
 \langle 0|
 (\sqrt{Q_{k'_1}}d^{}_{A_1k'_1} + \sqrt{1-Q_{k'_1}}d^{}_{A_2 k'_1})
\cdots  (\sqrt{Q_{k'_n}}d^{}_{A_1k'_n} + \sqrt{1-Q_{k'_n}}d^{}_{A_2 k'_n}).
\label{eq: rdm}
\end{align}

The above expression of the reduced density matrix looks tedious.
However, by introducing four matrices
\begin{align}
&[\rho_{11}]_i = (\sqrt{Q_i}d^{\dagger}_{A_1 i}+\sqrt{1-Q_i}d^{\dagger}_{A_2 i}) |0\rangle \langle 0| (\sqrt{Q_i}d^{}_{A_1 i}+\sqrt{1-Q_i}d^{}_{A_2 i}) 
= \left(\begin{array}{cccc} Q_i & \sqrt{Q_i(1-Q_i)} & 0 & 0 \\ \sqrt{Q_i(1-Q_i)} & 1- Q_i & 0 & 0 \\0 & 0 & 0 & 0 \\0 & 0 & 0 & 0\end{array}\right),  \notag\\
&[\rho_{00}]_i = |0 \rangle \langle 0|
 =\left(\begin{array}{cccc} 0 & 0 & 0 & 0 \\ 0 & 0 & 0 & 0 \\0 & 0 & 1 & 0 \\0 & 0 & 0 & 0\end{array}\right), \quad \notag\\
 &[\rho_{10}]_i =  (\sqrt{Q_i}d^{\dagger}_{A_1 i}+\sqrt{1-Q_i}d^{\dagger}_{A_2 i}) |0\rangle \langle 0| 
=  \left(\begin{array}{cccc} 0 &0& \sqrt{Q_i} & 0 \\ 0 & 0 &  \sqrt{1-Q_i}  & 0 \\0 & 0 & 0 & 0 \\0 & 0 & 0 & 0\end{array}\right), \notag\\
 & [\rho_{01}]_i =  |0\rangle \langle 0| (\sqrt{Q_i}d^{}_{A_1 i}+\sqrt{1-Q_i}d^{}_{A_2 i}) 
 =\left(\begin{array}{cccc} 0 &0& 0& 0 \\ 0 & 0 & 0  & 0 \\ \sqrt{Q_i}  &  \sqrt{1-Q_i} & 0  & 0  \\0 & 0 & 0 & 0\end{array}\right),
 \label{Eq: 4M}
\end{align}
the reduced density matrix can be expressed in terms of the Kronecker product of the combination of these four matrices.
For example
\begin{align}
&(\sqrt{Q_{k_1}}d^{\dagger}_{A_1k_1} + \sqrt{1-Q_{k_1}}d^{\dagger}_{A_2 k_1})
(\sqrt{Q_{k_2}}d^{\dagger}_{A_1k_2} + \sqrt{1-Q_{k_2}}d^{\dagger}_{A_2 k_2})
(\sqrt{Q_{k_3}}d^{\dagger}_{A_1k_3} + \sqrt{1-Q_{k_3}}d^{\dagger}_{A_2 k_3}) 
 |0 \rangle   \langle 0| \notag\\
&\times
 (\sqrt{Q_{k_1}}d^{}_{A_1k_1} + \sqrt{1-Q_{k_1}}d^{}_{A_2 k_1})
(\sqrt{Q_{k_2}}d^{}_{A_1k_2} + \sqrt{1-Q_{k_2}}d^{}_{A_2 k_2})   
(\sqrt{Q_{k_4}}d^{}_{A_1k_4} + \sqrt{1-Q_{k_4}}d^{}_{A_2 k_4}) \notag \\
=& [\rho_{11}]_{k_1}  \otimes [\rho_{11}]_{k_2} \otimes [\rho_{10}]_{k_3} \otimes [\rho_{01}]_{k_4} \otimes  [\rho_{00}]_{k_5} \cdots \otimes  [\rho_{00}]_{k_n}.
\end{align}

The partial transpose of these four matrices have the following forms
\begin{align}
&[\rho_{11}]^{T_{A_2}}_i = 
 \left(\begin{array}{cccc} Q_i & 0 & 0 & 0 \\0 & 1- Q_i & 0 & 0 \\0 & 0 & 0 & \sqrt{Q_i(1-Q_i)} \\0 & 0 & \sqrt{Q_i(1-Q_i)} & 0\end{array}\right),  \quad
[\rho_{00}]^{T_{A_2}}_i=
 \left(\begin{array}{cccc} 0 & 0 & 0 & 0 \\ 0 & 0 & 0 & 0 \\0 & 0 & 1 & 0 \\0 & 0 & 0 & 0\end{array}\right), \quad \notag\\
 &[\rho_{10}]^{T_{A_2}}_i = 
 \left(\begin{array}{cccc} 0 &0& \sqrt{Q_i} & 0 \\ 0 & 0 & 0  & 0 \\0 &  \sqrt{1-Q_i} & 0 & 0 \\0 & 0 & 0 & 0\end{array}\right), \quad
  [\rho_{01}]^{T_{A_2}}_i = 
 \left(\begin{array}{cccc} 0 &0& 0& 0 \\ 0 &  0  & \sqrt{1-Q_i}  & 0 \\ \sqrt{Q_i}  &0& 0  & 0 \\0 & 0 & 0 & 0\end{array}\right).
\end{align}
\end{widetext}
By summing all the combinations of all Kronecker product of these four partially transposed matrices,
the partially transposed reduced density matrix can be expressed as
\begin{align}
\rho_A^{T_{A_2}} = \sum_{n_i=11, 00,10,01} C_{n_1, n_2, \cdots ,n_N} \bigotimes_{i=1}^{N} [\rho_{n_i}]^{T_{A_2}}_i,
\label{Eq: general}
\end{align}
where $ C_{n_1, n_2, \cdots ,n_N} $ is the coefficient which can be computed from Eq. (\ref{eq: rdm}). 
The size of the partially transposed reduced density matrix is $(4^N )^2$, where $N$ is the total number of particles. 
The entanglement negativity is obtained by $\mathcal{E} = \ln \mathrm{Tr} |\rho_A^{T_{A_2}}|$.

\end{document}